\documentstyle[amsfonts,12pt]{article}
\def\one{1\hskip-.37em 1}

\def\half{\textstyle{\frac{1}{2}}}

\def\H{{\cal H}}
\def\D{{\cal D}}
\def\l{{\lambda}}

\def\ep{\epsilon}
\def\E{{\rm I}\hskip-.2em{\rm E}}
\def\ra{\rightarrow}
\def\tint{{\textstyle\int}}
\def\hg{{\hat g}}
\def\hp{{\hat\pi}}

\def\s{\hskip.08em}

\def\o{\overline}
\def\a{\alpha}
\def\b{\begin{eqnarray*}}     %takes no eqn numbers
\def\e{\end{eqnarray*}}       %takes no eqn numbers
\def\bn{\begin{eqnarray}}     %takes eqn numbers 
\def\en{\end{eqnarray}}       %takes eqn numbers
\def\<{\langle}
\def\>{\rangle}

\def\no{\nonumber}

\def\{{\lbrace}
\def\}{\rbrace}
\begin{document}
%\footnote{Electronic mail: klauder@phys.ufl.edu}
\title{The Utility of Coherent States and other \\Mathematical Methods in the\\Foundations of Affine Quantum Gravity\footnote{Presented at the X-International
Conference on Symmetry Methods in Physics, Yerevan, Armenia, August, 2003.}}
\author{John R. Klauder
\footnote{Electronic mail: klauder@phys.ufl.edu}\\
Departments of Physics and Mathematics\\
University of Florida\\
Gainesville, FL  32611}
%Email: klauder@phys.ufl.edu}
\date{}     %   Use   %\date{} to see the dates
\maketitle
\begin{abstract}
Affine quantum gravity involves (i) affine commutation relations
to ensure metric positivity, (ii) a regularized projection operator procedure
to accomodate first- and second-class quantum constraints, and (iii) a
hard-core interpretation of nonlinear interactions to understand and
potentially overcome nonrenormalizability. In this program, some of the 
less traditional mathematical methods employed are (i) coherent state 
representations, (ii) reproducing kernel Hilbert spaces, and (iii) 
functional integral representations involving a continuous-time
regularization. 
Of special importance is the profoundly different integration measure used 
for the Lagrange multiplier (shift and lapse) functions. These various concepts are first introduced on elementary systems to help motivate their application to affine quantum gravity.
\end{abstract}
%\vfill\eject
\section*{Introduction}
The unification of gravity and quantum theory offers a major challenge to
theoretical physics. The favored approaches of string theory and loop plus spin foam gravity use formulations that are in some sense rather far from the original classical theory of Einstein. Most workers feel this is necessary because of the usual difficulties encountered in quantizing gravity, namely, nonrenormalizability and anomalies, to mention just two. The program
of affine quantum gravity, which is relatively new \cite{aff}, attempts to stay closer to the standard classical theory so as to provide suitable touchstones along
the way. As a consequence it becomes necessary to deal directly
with some of the major problems, such as the two mentioned above. How one deals
with such difficult issues, and especially the role played by coherent
states in this effort, are part of the story told in this article.

As a pedagogical device we illustrate our basic methodology on simple quantum mechanical systems before we discuss the case of quantum gravity.

\subsection*{Coherent state basics}
It is well known that coherent states provide a useful bridge between a
classical theory and the corresponding quantum theory. Let us briefly recall how that
bridge works with a simple example. Let $Q$ and $P$ denote standard
Heisenberg self-adjoint operators satisfying the usual commutation
relation $[Q,P]=i\one$ with $\hbar=1$. Then we define canonical
coherent states by the relation
  \bn  |p,q\>\equiv e^{-iqP}\s e^{ipQ}\s|\eta\> \;, \en
where $|\eta\>$ denotes a normalized vector called the fiducial vector, 
which, in terms of the abbreviation 
  $ \<(\cdot)\>\equiv\<\eta|(\cdot)|\eta\>$,
is subject to the modest requirements that
  $\<\s Q\s\>=0$ and $\<\s P\s\>=0 $.
This condition on $|\eta\>$  has been referred to as ``physically centered''. Here, we add the additional
requirement that
  \bn \lim_{\hbar\ra0}\,\<(P^2+Q^2)\>=0\;, \label{e4} \en
a relation 
we refer to as ``physically attractive''. Given appropriate
domain conditions, it follows from (\ref{e4}) that
  \bn \lim_{\hbar\ra0}\<\s(P^2+Q^2)^m\s\>=0 \label{e5}\en
for arbitrary $m>0$. 
It is clear that the ground state of an harmonic oscillator satisfies these
conditions, but so do many other vectors as well.

If $\cal G$ denotes a quantum ``generator'' in a wide sense, then we assert
that 
   \bn G(p,q)\equiv\<p,q|\s{\cal G}\s|p,q\> \label{f6}\en
defines the ($\hbar$ augmented) classical generator $G(p,q)$ associated with $\cal G$. Of course,
this connection is not strictly what we usually mean by the classical generator since $\hbar$ has not been set equal to zero -- which explains the ``($\hbar$ augmented)'' phrase. In addition, we can also consider the expression
 \bn G_c(p,q)\equiv\lim_{\hbar\ra0}\,\<p,q|\s{\cal G}\s|p,q\>\;,  \label{f7}\en
which corresponds to the complete classical limit. The association between a quantum and classical generator illustrated by (\ref{f6}) and (\ref{f7}) is known as the ``weak correspondence principle'' \cite{wcp}.

To illustrate this situation, let us discuss a few
examples. For example, it follows that
  \bn \<p,q|\s Q\s|p,q\>=q\;, \hskip1cm \<p,q|\s P\s|p,q\>=p \;, \en
while
  \bn \<p,q|\s Q^2\s|p,q\>=q^2+\<\s Q^2\s\>\;, 
\hskip1cm \<p,q|P^2|p,q\>=p^2+\<P^2\>\;. \en
More generally, if ${\cal W}(P,Q)$ denotes an arbitrary polynomial, and subject to suitable domain conditions, then
 \bn W(p,q)\equiv\<p,q|\s{\cal W}(P,Q)\s|p,q\>={\cal W}(p,q)+{\cal O}(\hbar;p,q)\;,  \en
where under the condition (\ref{e5}), it follows that ${\cal O}(\hbar;p,q)\ra
0$ as $\hbar\ra0$.

A complete characterization of an operator is given in terms of its matrix elements. In particular, it is clear that
  \bn \<p,q|\s{\cal W}(P,Q)\s|p',q'\> \en
fully determines the operator in question.

\subsection*{Reproducing kernel Hilbert space}
By assumption the coherent states span the Hilbert space. Therefore a dense set of vectors may be written in the form
    \bn |\psi\>=\sum_{j=1}^J\s\alpha_j\s|p_j,q_j\>\;, \label{e7}\en
where $(p_j,q_j)\in{\mathbb R}^2$, $\alpha_j\in{\mathbb C}$, and $J<\infty$.
Another such vector may be taken as
  \bn |\phi\>=\sum_{k=1}^K\s\beta_k\s|p_{(k)},q_{(k)}\>\;,\hskip.6cm K<\infty\;.  \label{e8}\en
As {\it functional representatives} of these abstract vectors let us choose their inner product with arbitrary coherent states which leads to
\bn && \psi(p,q)\equiv \<p,q|\psi\>=\sum_{j=1}^J\s\alpha_j\s\<p,q|p_j,q_j\>\;,\\  && \phi(p,q)\equiv \<p,q|\phi\>=\sum_{k=1}^K\s\beta_k\s\<p,q|
p_{(k)},q_{(k)}\>\;.  \en
As the {\it inner product} between two such functional representatives we take
  \bn (\psi,\phi)\equiv\<\psi|\phi\>=\sum_{j,k=1}^{J,K}\s\alpha_j^*\s\beta_k\s
\<p_j,q_j|p_{(k)},q_{(k)}\>\;, \en
as follows from (\ref{e7}) and (\ref{e8}). We now have functional representatives, $\psi(p,q)$, $\phi(p,q)$, etc., and an inner product between them; all that
remains to make a Hilbert space is to complete the space by including the
limits of all Cauchy sequences in the norm $\|\psi\|\equiv +\sqrt{(\psi,\psi)}$. The result is the so-called (separable) reproducing kernel Hilbert space
in which the continuous function given by the coherent state overlap function
serves as the reproducing kernel. For additional information about such
Hilbert spaces see \cite{mesh}.

\section*{Affine Field Operators}
In a $3+1$ split of space and time, a subset of the basic kinematical operators chosen for affine quantum field theory involves the self-adjoint spatial metric $\hg_{ab}(x)$, where $a,b=1,2,3$. Moreover, we insist that the spectrum of the spatial metric is restricted so that $u^a\s\hg_{ab}(x)\s u^b>0$ for any set $\{u^a\}$ for which $\Sigma_a(u^a)^2>0$, a requirement that we call ``metric positivity''. To complete the set of basic kinematical operators we employ the ``momentric'' field $\hp^b_d(x)$. The
latter field is the self-adjoint operator associated with the classical momentric field $\pi^b_d(x)\equiv\pi^{bc}(x)\s g_{cd}(x)$, which involves both the ADM classical momentum $\pi^{bc}(x)$ and spatial metric 
$g_{cd}(x)$. Promoting the Poisson brackets satisfied by the fields $\pi^c_d$ and $g_{ab}$ to commutators leads us directly to the {\it affine commutation relations} (for $\hbar=1$) given by
   \bn 
  &&\hskip.2cm[\hp^a_b(x),\,\hp^c_d(y)]=\half\s i\s[\s\delta^c_b\hp^a_d(x)-\delta^a_d\hp^c_b(x)\s]\,\delta(x,y)\;,\no\\
  &&\hskip.1cm[\hg_{ab}(x),\,\hp^c_d(y)]=\half\s i\s[\s\delta^c_a\hg_{bd}(x)+\delta^c_b\hg_{ad}(x)\s]\,\delta(x,y)\;,\label{afc}\\
&&[\hg_{ab}(x),\,\hg_{cd}(y)]=0\;.   \no \en
Observe, that these relations define an infinite dimensional Lie algebra.
The reason for choosing these particular kinematical commutators follows 
directly from the fact that
\bn e^{i\tint \gamma^a_b(y)\s\hp^b_a(y)\,d^3\!y}\,\hg_{rs}(x)\,e^{-i\tint \gamma^a_b(y)\s\hp^b_a(y)\,d^3\!y}=(\s e^{\gamma(x)/2}\s)_r^{t}\,
\hg_{tu}(x)\,(\s e^{\gamma(x)/2}\s)_s^{u}\;. \en
The meaning of this relation is clear: Unitary transformations generated by the
self-adjoint momentric field manifestly preserve the desired spectral domain of
the spatial metric tensor ensuring that $u^a\s\hg_{ab}(x)\s u^b>0$ for any
set $\{u^a\}$ that is not identically zero.

\subsection*{Affine coherent states}
A representation of the basic affine operators is determined by a state on
the algebra they generate -- or equivalently, by the set of coherent states
  \bn |\pi,\gamma\>\equiv e^{i\tint\pi^{ab}(x)\s\hg_{ab}(x)\,d^3\!x}\,
e^{-i\tint\gamma^a_b(x)\s\hp^b_a(x)\,d^3\!x}\,|\eta\> \en
for all smooth functions $\pi^{ab}$ and $\gamma^a_b$ of compact support. Here
$|\eta\>$ is a suitable fiducial vector which, in effect, determines the
field operator representation. The appropriate (physical) restriction on
this operator representation arises from fully enforcing the gravitational
constraints. However, according to Dirac \cite{dir}, quantization should be
carried out {\it first} while the introduction of constraints should come      {\it second}.
Thus, we are obliged to choose an initial -- and temporary -- field 
operator representation just to get started. To this end it proves 
convenient to choose $|\eta\>$ so that 
  \bn &&\<\pi'',\gamma''|\pi',\gamma'\>\equiv\<\pi'',g''|\pi',g'\>=\exp\!
\bigg[-\int b(x)\,d^3\!x \no\\
  &&\hskip.6cm\times\ln\!\bigg(\frac{\det\{\half[g''^{ab}(x)+g'^{ab}(x)]+ib(x)^{-1}[\pi''^{ab}(x)-\pi'^{ab}(x)]\s\}}{\{\s\det[g''^{ab}(x)]\,\det[g'^{ab}(x)]\s\}^{1/2}}\s\bigg)\s\bigg]\!.  \label{e12}\en
Here, the symmetry of $|\eta\>$ is such that instead of all nine components of $\gamma$, the coherent states depend on only the six components of $g$, which
are defined by
   \bn  g_{ab}(x)\equiv (\s e^{\gamma(x)/2}\s)_a^{c}\,{\tilde g}_{cd}(x)\,(\s e^{\gamma(x)/2}\s)_b^{d}\;,  \en
where 
  \bn {\tilde g}_{ab}(x)\equiv\<\eta|\s \hg_{ab}(x)\s|\eta\>  \en
arises as a property of $|\eta\>$.
The scalar density $b(x)$, $0< b(x)< \infty$, arises as a property of $|\eta\>$ as well. As usual, $g^{ab}(x)$ is the inverse metric defined by 
$g^{ab}(x)\s g_{bc}(x)=\delta^a_c$ for each $x$.

Observe that the coherent state overlap (\ref{e12}) is a jointly
continuous function of its arguments, e.g., in the topology of the 
test function space ${\cal D}$. 

\subsection*{Reproducing kernel Hilbert space}
Just as in the elementary example, we can use the coherent state overlap function
(\ref{e12}) as a reproducing kernel to construct a reproducing kernel Hilbert space. In particular, functional representatives in a dense set of the Hilbert space may be given by 
 \bn && \psi(\pi,g)\equiv\sum_{j=1}^J\s\alpha_j\s\<\pi,g|\pi_j,g_j\>\;,\hskip.5cm J<\infty\;,\\
&& \phi(\pi,g)\equiv\sum_{k=1}^K\s\beta_j\s\<\pi,g|\pi_{(k)},g_{(k)}\>\;, \hskip.5cm K<\infty\;,\en
etc. As an inner product between two such vectors we choose
 \bn (\psi,\phi)\equiv\sum_{j,k=1}^{J,K}\s\alpha_j^*\s\beta_k
\s\<\pi_j,g_j|\pi_{(k)},g_{(k)}\>\;.  \en
We may complete this Hilbert space by introducing all limit elements of
Cauchy sequences in the norm $\|\psi\|\equiv+\sqrt{(\psi,\psi)}$, in
complete analogy to what we did in the elementary example. The result is
the separable reproducing kernel Hilbert space with (\ref{e12}) serving as the
reproducing kernel.

\section*{Imposition of Constraints}
To explain our procedure for the imposition of constraints we return to an
$N$ degree-of-freedom model, $N<\infty$. Let us suppose there are 
classical constraints for this model given by the 
conditions $\phi_\a(p,q)=0$ for
$1\le\a\le A$, where $p=(p^1,p^2,\ldots,p^N)$ and $q=(q^1,q^2,\ldots,q^N)$. 
Upon quantization, these constraints become self-adjoint
operators $\Phi_\a(P,Q)$, $1\le\a\le A$. Ideally, there should be a nonvanishing subspace ${\frak H}_{phys}$ of the original Hilbert space ${\frak H}$ for
which $\Phi_\a\s|\psi\>_{phys}=0$ for all $|\psi\>_{phys}\in{\frak H}_{phys}$
\cite{dir}. Unfortunately, this ideal situation does not always occur. As
a replacement for this criterion we introduce a {\it projection operator} 
  \bn \E=\E(\Sigma\s\Phi_\a^2\le\delta(\hbar)^2)\;, \en
i.e., a projection operator such that
 \bn 0\le \E\s(\Sigma\Phi_\a^2)\s\E\le\delta(\hbar)^2\s\one\;,  \en
where $\delta(\hbar)$ is a regularization parameter. We define ${\frak H}_{phys}=\E\s{\frak H}$ as the regularized physical Hilbert space. The general idea is to reduce the regularization parameter $\delta(\hbar)$ to an appropriate value for each situation. For example, if $\Sigma\s\Phi_a^2=J_1^2+J_2^2+J_3^2$, the Casimir operator for SU(2), then $\delta(\hbar)^2 =\half\hbar^2$ (or any $\delta(\hbar)$, $0\le\delta(\hbar)^2<(3/4)\hbar^2)$ is sufficient to ensure that $\Sigma\s J_k^2=0$. If $\Sigma\s\Phi_a^2=P^2+Q^2$, then $\delta(\hbar)^2=\hbar$ (or any $\delta(\hbar)$, $\hbar\le\delta(\hbar)^2<3\hbar$) ensures that $\E$ projects unto vectors $|\psi\>_{phys}$ for which $(Q+iP)\s|\psi\>_{phys}=0$. This procedure enables
a consideration of first- and second-class constraints within the same formulation without any need either to eliminate the second-class constraints before quantization or to introduce Dirac brackets for them. Other constraint situations can also be treated; see \cite{kla7}.

It is noteworthy that for any set of constraint operators, there exists a universal integral representation to construct the projection operator \cite{uni}. In
particular, it follows that
  \bn \E(\Sigma\s\Phi_a^2\le\delta(\hbar)^2)\equiv \int{\sf T}\s e^{-i\tint_{t_1}^{t_2}\lambda^\a(t)\s\Phi_\a\,dt}\,\D R(\lambda)\;,  \label{g26}\en
where ${\sf T}$ denotes time ordering, $R(\lambda)$ is a formal measure
on $c$-number functions $\{\lambda^\a(t)\}$, and $t_2-t_1$ corresponds to any positive time interval. The  integral in (\ref{g26}) is constructed in a two-step procedure. First, a Gaussian functional integral leads to
  \bn e^{i\gamma\s(t_2-t_1)\s\Sigma\s\Phi_\a^2}={\cal N}\int {\sf T}\s
e^{-i\tint \lambda^\a(t)\s\Phi_\a\,dt -(i/4\gamma)\tint\s\Sigma\s\lambda^\a(t)^2\,dt}
\,\Pi\s\D \lambda^\a\;.  \en
Second, we integrate over $\gamma$ as follows:
 \bn  \lim_{\zeta\ra0^+}\int \frac{\sin[\gamma\s(t_2-t_1)\s(\delta(\hbar)^2+\zeta)]}
{\pi\s\gamma}\,e^{i\gamma\s(t_2-t_1)\s\Sigma\s\Phi_\a^2}\,d\gamma=\E(\Sigma\s\Phi_\a^2\le\delta(\hbar)^2)\;.  \en

The integral representation (\ref{g26}) may be used within a coherent state path integral representation of the propagator. We focus on \cite{kla7} 
\bn &&\<p'',q''|\s\E\s e^{-i(\E\H\E)T}\s\E\s|p',q'\> \no\\
  &&\hskip.5cm=\lim_{\ep\ra0}\,\<p'',q''|e^{-i\H\ep}\E e^{-i\H\ep}\E\cdots e^{-i\H\ep}\E|
p',q'\>\;, \en
where there are $(L+1)$ short-time evolution operators $e^{-i\H\ep}$, and
$(L+1)\ep=T$. Insertion of $L$ coherent state resolutions of unity leads to
  \bn \lim_{\ep\ra0}\int\!\cdots\!\int\prod_{l=0}^L\<p_{l+1},q_{l+1}|e^{-i\H\ep}\E|p_l,q_l\>\,
\prod_{l=1}^L\,d^N\!p_l\,d^N\!q_l/(2\pi)^N\;,  \en
where $p_{L+1},q_{L+1}=p'',q''$ and $p_0,q_0=p',q'$. In turn, it follows that
 \bn &&\<p'',q''|\s\E\s e^{-i(\E\H\E)T}\s\E\s|p',q'\>\no\\
 &&\hskip.5cm=\lim_{\ep\ra0}\int\!\cdots\!\int\prod_{l=0}^L\<p_{l+1},q_{l+1}|
e^{-i\ep(\H+\lambda_l^\a\Phi_\a)}|p_l,q_l\>\no\\
&&\hskip3.5cm\times\prod_{l=1}^L \,d^N\!p_l\,
d^N\!q_l/(2\pi)^N\,\D R(\lambda_l) \no\\
 &&\hskip.5cm={\cal M}\int e^{i\tint[p\cdot{\dot q}-H(p,q)-\lambda^\a(t)\phi_\a(p,q)]\,dt}\,\D p\,\D q\,\D R(\lambda)\;, \en
where $H(p,q)\equiv\<p,q|\H|p,q\>$ and $\phi_\a(p,q)\equiv\<p,q|\Phi_\a|p,q\>$. 

In this fashion we see how repeated insertions of the projection operator lead to temporal evolution entirely within the physical Hilbert space. Moreover,
we see how this evolution can be realized by a suitably interpreted path integral which does not involve the usual flat measure on the Lagrange multipliers but, instead, uses the measure $R(\lambda)$ that is designed to enforce the quantum constraints rather than the classical constraints. 

Although this is not the only way the integral representation for the projection operator (\ref{g26}) can be used to formulate a path integral, it is probably the most straightforward construction and readily illustrates the basic principles involved.

\subsection*{Constraints in quantum gravity}
For quantum gravity, there are four constraint fields, which, from a classical
point of view, comprise an open first-class system. The quantum constraints, however, exhibit an anomaly, and thus they are partially second class in nature. In particular, the diffeomorphism ($\H_a,\;a=1,2,3)$,  and Hamiltonian ($\H$) constraint
operator fields fulfill the commutation relations 
\bn  && [\H_a(x),\H_b(y)]=i\s[\delta_{,a}(x,y)\,\H_b(x)-\delta_{,b}(x,y)\,\H_a(x)]\;, \\
    && \hskip.18cm[\H_a(x),\H(y)]=i\s\delta_{,a}(x,y)\,\H(x)\;,  \\
 && \hskip.36cm[\H(x),\H(y)]= i\s\half\s\delta_{,a}(x,y)[\,\hg^{ab}(x)\,\H_b(x)+\H_b(x)\,\hg^{ab}(x)\,]\;.  \label{e31}\en
Ideally, one asks that $\H_a(x)\s|\psi\>_{phys}=0$ as well as $\H(x)\s|\psi\>_{phys}=0$ for all $a$ and $x$, and for all $|\psi\>_{phys}\in{\frak H}_{phys}$. However, this ideal situation is not possible because it is almost surely the case that $\hg^{ab}(x)\s|\psi\>_{phys}\not\in{\frak H}_{phys}$, and therefore it does not follow that $[\H(x),\H(y)]\s|\psi\>_{phys}=0 $ as would be required.  This inconsistency of Eq.~(\ref{e31}) gives rise to the gravitational anomaly -- a partially second-class behavior --  
that makes conventional 
treatments of quantum gravity especially difficult. However, as noted above
the projection operator method treats first- and second-class constraints
in the very same manner, the only difference being how small the
regularization parameter $\delta(\hbar)^2$ can be made.
  
By introducing a cutoff to regularize the quantum constraints, we can imagine constructing a
projection operator $\E$ onto a regularized physical Hilbert space in which the regularized quantum constraints are fulfilled to a certain degree. Such a cutoff can be 
introduced in a variety of ways, and for simplicity we will leave this 
necessary cutoff implicit. At a later point in the calculation it would be
necessary to remove this cutoff as well, but we will not examine this
important issue either. Instead, we go straight to the heart of the matter and note that
there is a functional integral representation \cite{aff} for the coherent-state matrix elements of the projection operator onto the regularized physical Hilbert space  given by
\bn  && \<\pi'',g''|\s\E\s|\pi',g'\>  \no\\
  &&\hskip1cm=\int \<\pi'',g''|{\bf T}\,e^{-i\tint[\s N^a\s\H_a+N\s\H\s]\,d^3\!x\,dt}\s|\pi',g'\>\,\D R(N^a,N)\no\\
 &&\hskip1cm=\lim_{\nu\ra\infty}{\o{\cal N}}_\nu\s\int e^{-i\tint[g_{ab}{\dot\pi}^{ab}+N^aH_a+NH]\,d^3\!x\,dt}\no\\
  &&\hskip1.5cm\times\exp\{-(1/2\nu)\tint[b(x)^{-1}g_{ab}g_{cd}{\dot\pi}^{bc}{\dot\pi}^{da}+b(x)g^{ab}g^{cd}{\dot g}_{bc}{\dot g}_{da}]\,d^3\!x\,dt\}\no\\
  &&\hskip2cm\times\bigg[\Pi_{x,t}\,\Pi_{a\le b}\,d\pi^{ab}(x,t)\,dg_{ab}(x,t)\bigg]\,\D R(N^a,N)\;. \label{f39}\en
Implicit in these expressions are cutoffs in the constraint operators $\H_a$ and $\H$, and correspondingly in the $c$-number symbols $H_a$ and $H$ that arise from the constraint operators and which appear in the functional integral as
their representatives.

Note the appearance of the measure $\D R(N^a,N)$ on the Lagrange multiplier fields, $N_a$ (the shift) and $N$ (the lapse). It is this measure, in contrast to the usual flat measure on these fields, which leads to the projection operator $\E$ that projects the original Hilbert space $\frak H$ onto the regularized physical Hilbert space ${\frak H}_{phys}$. 

Note also the appearance of a limit as $\nu\ra\infty$ as well as a $\nu$ dependent factor in the integrand. This factor and the limit are connected with a different kind of regularization of the functional integral 
that may be used instead of 
the usual lattice regularization. The indicated form represents 
a {\it continuous-time
regularization} which involves a Wiener-like measure that controls
the nature of the paths. The result of interest arises in the ultra-diffusive limit in which the diffusion constant $\nu$ diverges. Continuous-time regularization procedures have been well studied for phase-space path integrals 
approriate to canonical, spin, and affine variables \cite{daub}, and they have the virtue that they {\it automatically} lead to a quantum mechanical 
representation of the corresponding
expression in terms of coherent states with a fiducial vector given by an extremal weight vector. Recently, additional studies of such path integral representations have been made in the case of weak coherent states for the affine group 
when a traditional resolution of unity as a local integral fails to exist 
\cite{hart}.

Let us add that the quantity $\<\pi'',g''|\s\E\s|\pi',g'\>$ may also be used
as a reproducing kernel to build the reproducing kernel Hilbert space associated with the regularized physical Hilbert space ${\frak H}_{phys}$ in exactly the
same way that the original coherent state overlap $\<\pi'',g''|\pi',g'\>$ 
may be used to build the reproducing kernel Hilbert space associated with the 
original Hilbert space $\frak H$. 

Equation (\ref{f39}) represents as far as we can presently go in our formal development. Despite the canonical appearance of ({\ref{f39}), we emphasize that this functional integral representation has been based on the affine commutation relations (\ref{afc}) and {\it not} on any canonical commutation relations. 

\section*{Hard-Core Interactions in Quantum \\Mechanics}
Let us again return to the world of quantum mechanics to motivate the
next issue of concern. Consider an imaginary-time path integral 
for a single degree-of-freedom problem formally given by
  \bn I(\l)\equiv{\cal N}\int \exp\{-\half\tint[\s{\dot x}(t)^2+m^2\s x(t)^2]\,dt-\l\tint x(t)^{-4}\,dt\s\}\,\D x\;, \label{u7} \en
where the path integral runs over continuous paths for which $x(0)=x'$ and $x(T)=x''$, namely, all paths are pinned at the initial and final times, $t=0$ and $t=T$, respectively. This example clearly pertains to an oscillator with
a singular potential and a coupling constant $\l$ that we require to be nonnegative, $\l\ge0$. To help interpret (\ref{u7}), it proves useful to first {\it regularize} the singularity of the inverse quartic interaction. However, no matter how one attempts to regularize the singularity of the inverse quartic interaction, so as to give unambiguous meaning to the path integral, and subsequently proceeds to remove that regularization, it is known that the result leads to a {\it discontinuous perturbation} of the oscillator \cite{kbook}. A discontinuous perturbation has the property that it leaves an indelible imprint on the original system such that once
the interaction is introduced {\it it cannot be completely removed} as the coupling constant $\l\ra0$. In other words, 
  \bn \lim_{\l\ra0}\s I(\l)=I'(0)\not=I(0)\;.  \en
Singular interactions with this property are called 
{\it hard-core interactions}.
What happens is the following. Whenever $\l>0$, those paths
allowed by the free action that reach or cross the point of singularity, 
$x=0$, lead to a divergent value for the 
interaction term, with the property that
the paths in question are all {\it projected out} of the integration for any positive value of $\l$, however small. Thus those paths make no contribution to the path integral $I(\l)$ for any $\l>0$, and, as a consequence, 
as $\l\ra0$, those
paths never reappear, and the result $I'(0)$ is based on the oscillator but
has a contribution from {\it only} those continuous paths $x(t)$ for which 
$x(t)\not=0$ for all $t$ such that $0\le t\le T$. The evaluation of the
resultant path integral with the restricted set of paths defines
the expression $I'(0)$ and it clearly gives rise to a different result than
if the interaction had never been present in the first place, namely, $I(0)$, 
which corresponds to the free theory, i.e., the usual imaginary-time 
oscillator. The theory 
implictly defined by $I'(0)$ is called the ``pseudofree theory''. 

\subsection*{Hard-core interactions in field theory}
The kind of behavior illustrated above is not limited to the inverse 
quartic interaction but arises for any interaction of the form 
$|x|^{-\a}$ whenever $\a>2$. There are good reasons to make the analogy 
of such discontinuous perturbations with nonrenormalizable interactions
as they are known in quantum field theory. The full story of this analogy
is presented in Chap.~8 in \cite{kbook}. In other words, it 
is reasonable to suppose that what are regarded as nonrenormalizable interactions in quantum field theory behave as they do because they are in fact discontinuous perturbations that act as hard cores within appropriate functional integral formulations. Moreover, certain specialized nonrenormalizable models exhibit exactly the stated behavior; see, e.g., Chaps.~9 \& 10 in \cite{kbook}. These models possess enough symmetry so that solutions can be constructed outside of perturbation theory on the basis of generally accepted principles. Based on the experience gained with such models, it is our strong conviction that all nonrenormalizable quantum field theories can be understood as discontinuous perturbations that act as hard cores within functional integrals. 

Of course, there is an important difference in the nature of the 
excluded paths between what happens in the quantum mechanical case 
and the field theory case. For quantum mechanics,
the interactions exhibit singularities at finite positions (e.g., $x=0$, as
is the case for the interaction $x^{-4}$), while for the field case, the interactions exhibit singularities for fields that themselves have singular behavior
at some point in Euclidean spacetime, e.g., a field having the distributional behavior $|x|^{-\gamma}$ near $x=0$, where $\gamma$ is chosen such 
that this local behavior is acceptable for the free term but unacceptable 
for the interaction term; see Chap.~8 in \cite{kbook}. 

For quantum mechanical cases it is quite straightforward to identify
which paths should be excluded and which paths are to be retained. On the
other hand, in the case of quantum fields the situation is far more difficult. 
In addition, it is one thing to say that functions with certain singular behavior are
to be excluded, but it is a far more difficult thing to say how, in fact, 
{\it operationally} to accomplish that exclusion. For  covariant, 
nonrenormalizable
scalar fields, a proposal has recently been put forward \cite{phi4} that
identifies a {\it novel, nonclassical ($\propto\hbar^2$) counterterm}, which, 
it is conjectured, captures the
effect of the hard-core character of the interaction, a counterterm that
remains behind -- as any hard-core portion of an interaction must certainly
do -- even after the strength of the interaction is reduced to zero. The
proposal offered is at a stage where Monte Carlo computer studies could
illuminate this proposal to a considerable degree; unfortunately, such
computer studies have yet to be made. If such computer studies were made, however, and they
confirmed that the hard-core picture makes good sense and also led to
nontrivial results for such nonrenormalizable models as $\phi^4_n$, for
spacetime dimensions $n\ge5$, then we would have greater confidence in
their possible utility in the study of quantum gravity. Since we do not yet have this
additional degree of support, we are obliged to rely on the conjecture
that the nonrenormalizable aspect of traditional quantum gravity can be
understood -- and eventually dealt with -- by invoking the hard-core 
hypothesis, even if at this stage we do not fully know how to actually 
realize this proposal.

At any rate, we can make a few reasonable conjectures as to how the 
appearance of the hard-core terms may enter. Just as with covariant
scalar fields, we expect the counterterm(s) to be {\it atypical} and {\it not}
what would be predicted on the basis of perturbation theory. After all,
perturbation theory is based on the {\it assumption} that the interacting
theory is continuously connected to the noninteracting theory, indeed, explicitly in such a way that as the coupling constant goes to zero, one passes
continuously from the interacting theory to the noninteracting one.
But, for hard-core interactions, that is {\it exactly what cannot happen}.
Thus we are led to expect modifications of the constraint
operators $\H_a$ and $\H$, which will then lead to $O(\hbar)$ modifications
to their $c$-number symbols $H_a$ and $H$ that enter into the functional integral (\ref{f39}). Since these all-important modifications are unknown at
present, we are not yet in a position to try to use (\ref{f39}) in order
to evaluate, even approximately, the coherent state matrix element of the
projection operator, $\<\pi'',g''|\s\E \s|\pi',g'\>$. 

In conclusion, we
expect the next level of understanding in this program to arise from the study of (i) $\phi^4_n$, $n\ge5$, models and (ii) simple models with anomalous constraint behavior. However, predicting the future is known to be fairly risky!

\end{document}